\documentclass[10pt,letterpaper]{article}
\usepackage[top=0.85in,left=2.75in,footskip=0.75in,marginparwidth=2in]{geometry}
\usepackage{amsmath}
\usepackage{amsfonts}
\usepackage{booktabs}
\usepackage[utf8x]{inputenc}

\usepackage{cite}

\usepackage{nameref,hyperref}

\usepackage[right]{lineno}

\usepackage{microtype}
\DisableLigatures[f]{encoding = *, family = * }

\raggedright
\usepackage{changepage}

\usepackage[aboveskip=1pt,labelfont=bf,labelsep=period,singlelinecheck=off]{caption}

\makeatletter
\renewcommand{\@biblabel}[1]{\quad#1.}
\makeatother

\usepackage{lastpage,fancyhdr,graphicx}
\usepackage{epstopdf}
\fancyheadoffset[L]{2.25in}
\fancyfootoffset[L]{2.25in}

\usepackage{color}

\definecolor{Gray}{gray}{.25}

\usepackage{graphicx}

\usepackage{sidecap}

\usepackage{wrapfig}
\usepackage[pscoord]{eso-pic}
\usepackage[fulladjust]{marginnote}
\reversemarginpar

\begin{document}
\vspace*{0.35in}

\begin{flushleft}
{\Large
\textbf\newline{DeepGO: Predicting protein functions from
  sequence and interactions using a deep ontology-aware classifier}
}
\newline
\\
Maxat Kulmanov\textsuperscript{1},
Mohammed Asif Khan\textsuperscript{1},
Robert Hoehndorf\textsuperscript{1, *},
\\
\bigskip
\bf{1} Computational Bioscience Research Center, \\
  Computer, Electrical and Mathematical Sciences \& Engineering
  Division,\\ King Abdullah University of Science and Technology,\\
  4700 King Abdullah University of Science and Technology, \\ Thuwal
  23955-6900, Kingdom of Saudi Arabia
\\
\bigskip
* robert.hoehndorf@kaust.edu.sa

\end{flushleft}

\section*{Abstract}
\textbf{Motivation:} A large number of protein sequences are
  becoming available through the application of novel high-throughput
  sequencing technologies. Experimental functional characterization of
  these proteins is time-consuming and expensive, and is often only
  done rigorously for few selected model organisms. Computational
  function prediction approaches have been suggested to fill this
  gap. The functions of proteins are classified using the Gene
  Ontology (GO), which contains over 40,000 classes. Additionally,
  proteins have multiple functions, making
  function prediction a large-scale, multi-class, multi-label problem. \\
  \textbf{Results:} We have developed a novel method to predict
  protein function from sequence. We use deep learning to learn
  features from protein sequences as well as a cross-species
  protein-protein interaction network. Our approach specifically
  outputs information in the structure of the GO and utilizes the
  dependencies between GO classes as background information to
  construct a deep learning model. We evaluate our method using the
  standards established by the Computational Assessment of Function
  Annotation (CAFA) and demonstrate a significant improvement over
  baseline methods such as BLAST, with significant improvement for
  predicting cellular locations.  \\
  \textbf{Availability and Implementation:}
  Web server: \url{http://deepgo.bio2vec.net}, Source code:
  \url{https://github.com/bio-ontology-research-group/deepgo}

\section{Introduction}

Advances in sequencing technology have led to a large and rapidly
increasing amount of genetic and protein sequences, and the amount if
expected to increase further through sequencing of additional
organisms as well as metagenomics.  Although knowledge of protein
sequences is useful for many applications, such as phylogenetics and
evolutionary biology, understanding the behavior of biological systems
additionally requires knowledge of the proteins' functions.
Identifying protein functions is challenging and commonly requires
{\em in vitro} or {\em in vivo} experiments \cite{Costanzoaaf1420},
and it is obvious that experimental functional annotation of proteins
will not scale with the amount of novel protein sequences becoming
available.

One approach to address the challenge of identifying proteins'
functions is the computational prediction of protein functions
\cite{cafa}.  Function prediction can use several sources of
information, including protein-protein interactions \cite{Sharan88},
genetic interactions \cite{Costanzoaaf1420}, evolutionary relations
\cite{paint}, protein structures and structure prediction methods
\cite{Konc2013}, literature \cite{Verspoor2014}, or combinations of
these \cite{SokolovB10}.  These methods have been developed for many
years, and their predictive performance is improving steadily
\cite{cafa}.

There are several key challenges for protein function prediction
methods. One of these is the complex relation between protein
sequence, structure and function \cite{molecularcell}; despite
significant progress in the past years in protein structure prediction
\cite{casp}, it still requires large efforts to predict protein
structure with sufficient quality to be useful in function
prediction. Another challenge is the large and complex output space
for any classification method. Protein functions are classified using
the Gene Ontology (GO) \cite{Ashburner2000} which contains over 40,000
functions and cellular locations. Additionally, the GO contains
strong, formally defined relations between functions that need to be
taken into account during function prediction to ensure that these
predictions are consistent \cite{SokolovB10, cafa}. The formal
dependencies between classes in GO also lead to the situation where
proteins are assigned to multiple function classes in GO, for
different levels of abstraction. Furthermore, several proteins do not
only have a single function but may be peiotropic and have multiple
different functions, making function prediction inherently a
multi-label, multi-class problem. A final challenge is that proteins
do not function in isolation. In particular higher-level physiological
functions that go beyond simple molecular interactions, such as {\em
  apoptosis} or {\em regulation of heart rate}, will require other
proteins and cannot usually be predicted by considering a single
protein in isolation.  Due to these challenges, it is also not obvious
what kind of features should be used to predict the functions of a
protein, and whether they can be generated efficiently for a large
number of proteins.

Here, we present a novel method for predicting protein functions from
protein sequence and known interactions. We combine two forms of
representation learning based on multiple layers of neural networks to
learn features that are useful for predicting protein functions, one
method that learns features from protein sequence and another that
learns representations of proteins based on their location in an
interaction network. We then utilize these features in a novel deep
neuro-symbolic model that is built to resemble the structure and
dependencies between classes that exist within the GO, refine
predictions and features on each level of GO, and ultimately optimize
the performance of function prediction based on the performance over
the whole ontology hierarchy. 

We demonstrate that our model improves performance of function
prediction over a BLAST baseline, and performs particularly well in
predicting cellular locations of proteins. The main advantage of our
approach is that it does not rely on manually crafted features but is
entirely data-driven.


\section{Materials and Methods}

\subsection{Datasets}
For our experiments, we use the Gene Ontology (GO)
\cite{Ashburner2000}, downloaded on 05 January 2016 from
\url{http://geneontology.org/page/download-ontology} in OBO
format. The version of GO has 44,683 classes of which 1,968 are
obsolete. GO has three major branches, one for biological processes
(BP), molecular functions (MF) and cellular components (CC), each
containing 28,647, 10,161, and 3,907 classes, respectively.

We use SwissProt's \cite{Boutet2016} reviewed and manually annotated
protein sequences with GO annotations downloaded on 05 January 2016
from \url{http://www.uniprot.org/uniprot/}. The dataset contains
553,232 proteins, and 525,931 proteins have function
annotations. Furthermore, we select proteins with annotations with
experimental evidence code (EXP, IDA, IPI, IMP, IGI, IEP, TAS and IC)
and filter the proteins by maximum length of 1,002 ignoring proteins
with ambiguous amino acid codes (B, O, J, U, X, Z) in their
sequence. Our final dataset contains 60,710 proteins annotated with
27,760 classes (19,181 in BP, 6,221 in MF, and 2,358 in CC).

\subsection{Training}


We trained three models, one for each sub-ontology in GO. First, we
propagate annotations using the GO ontology structure and randomly
split proteins into a training set (80\%) and testing set (20\%). Due
to computational limitations and the small number of annotations for
very specific GO classes, we ranked GO classes by their number of
annotations and selected the top 932 terms for BP, 589 terms for MF
and 436 terms for the CC ontology. These cutoff values correspond to
selecting only classes with the minimum number of annotations
250, 50, and 50, for BP, MF, and CC, respectively.

We create three binary label vectors for each protein sequence, one
for each of the GO hierarchies. If a protein sequence is annotated
with a GO class from our lists of selected classes, then we assign $1$
to the term's position in the binary label vector and use it as
positive sample for this term.  Otherwise, we assign $0$ and use it as
negative sample.  For training and testing, we use proteins which have
been annotated with at least one GO term from the set of the GO terms
for the model.

\subsection{Data Representation}
The input of our model is the amino acid (AA) sequence of a
protein. Each protein is a character sequence composed of 20 unique AA
codes. We generate trigrams of AA from the protein sequence. The
trigrams can be represented as one-hot encoding vectors of length
8,000; however, the sparse nature of one-hot encodings only provides a
limited generalization performance. To address this limitation, we use
the notion of dense embeddings \cite{hinton1986learning,
  bengio2003neural}.  An embedding is a lookup table used for mapping
each code in a vocabulary to a dense vector. Initially, we initialized
the vectors randomly and then learn the actual vector-based
representations as an additional layer in our network architecture
during training. This approach allows us to learn meaningful vectors,
i.e., vectors that resemble correlations within the data that can be
utilized as features to predict protein functions.  We have also
performed experiments (on a smaller dataset) with one-hot encodings of
AA trigrams, and found that dense representation performs better than
one-hot encoding.

We built a vocabulary of unique AA trigrams where each trigram is
represented by its 1-based index. Using this vocabulary, we encoded a
sequence of length $1002$ as a vector of $1000$ indices. If the length
of the sequence is less than $1002$, we pad the vector with zeros. We
ignore all the proteins with sequence length more than $1002$. The
first layer in the deep learning model is intended to learn embeddings
where each index is mapped to a dense vector by referring to a lookup
table, using an embedding size of $128$ and therefore representing a
protein sequence of length of $1002$ as a matrix of $1000 \times 128$.  

\subsection{Convolutional Neural Network}
Convolutional Neural Networks (CNNs) are biologically inspired NN
which try to mimic the receptive field of biological neuron. In CNNs,
convolution operations are applied over the input layer to compute the
output \cite{Bengio95convolutionalnetworks}. They exploit local
correlation by enforcing local connections between neurons of adjacent
layers, where each region of the input is connected to a neuron in the
output. Having multiple convolution filters helps in learning multiple
features and providing insights into multiple facets of the data. In
our work, we used 1-dimensional (1D) convolution over protein sequence
data. The 1D convolution exploits sequential correlation. If we have
an input $g(x) \in [ 1,l] \rightarrow \mathbb{R}$ and a kernel
function $f(x) \in [1,k] \rightarrow \mathbb{R}$, the convolution
$h(y)$ between $f(x)$ and $g(x)$ with stride $d$ is defined as:
\begin{equation}
h(y) = \sum\limits_{x=1}^k f(x)\cdot g(y \cdot d-x+c)
\end{equation}
where $c= k-d+1$ is an offset constant. The output $h_j(y)$ is
obtained by a sum over $i$ of the convolutions between $g_i(x)$ and
$f_{ij}(x)$. The output vector $h$ represents the feature map learned
through convolution.

The resulting feature map will contain redundant information and is of
significant size. Therefore, to reduce the feature space, redundant
information is discarded through temporal max-pooling
\cite{collobert2011natural}.  This operation selects the maximum value
over a window of some length $w$. The features after convolution and
the temporal pooling layer are intended to be higher level
representation of protein sequences which can then be used as input to
fully connected layers for classification.

For our experiments, we used one 1D Convolution layer with $32$
filters of size $128$ which are applied on the embedding matrix of
each sequence, and a 1D max-pooling layer with pool length of $64$ and
stride of $32$. Each filter is intended to learn a specific type of
feature, and multiple filters may enable learning of different aspects
of the underlying data. The output of the 1D max-pooling layer is a
vector with length of $832$.

\subsection{Protein-protein interaction (PPI) network features}
In addition to protein sequences, we use protein-protein interaction
(PPI) networks for multiple species from the STRING database
\cite{stringdb10}, filtered by confidence score of 300 and connected
with orthology relations from the EggNOG database \cite{eggnog2016} by
creating a symmetric {\em ortholog-of} edge for each orthology
group. To further separate proteins by the orthology group to which
they belong, we introduce a new orthology relation for each orthology
group in eggNOG.  In total, the network consists of 8,478,935
proteins, 190,649 edge types and 11,586,695,610 edges. Using this
heterogeneous network, we generated knowledge graph embeddings of size
$256$ for each protein \cite{alshahrani16}.

Since our model is based on UniProt protein identifiers, we mapped
nodes in the network to UniProt identifiers using the identifier
mapping provided by STRING. We mapped 6,960,395 proteins in UniProt to
our network and the resulting knowledge graph embeddings. For the
proteins with missing network representations, we assigned a vector of
zeros. We combined the knowledge graph embeddings for the nodes with
the output of the max-pooling layer of length 832 as a combined
feature vector.

\subsection{Hierarchical classification layout}

\begin{figure*}[!ht]
  \centering
  \includegraphics[width=0.80\textwidth]{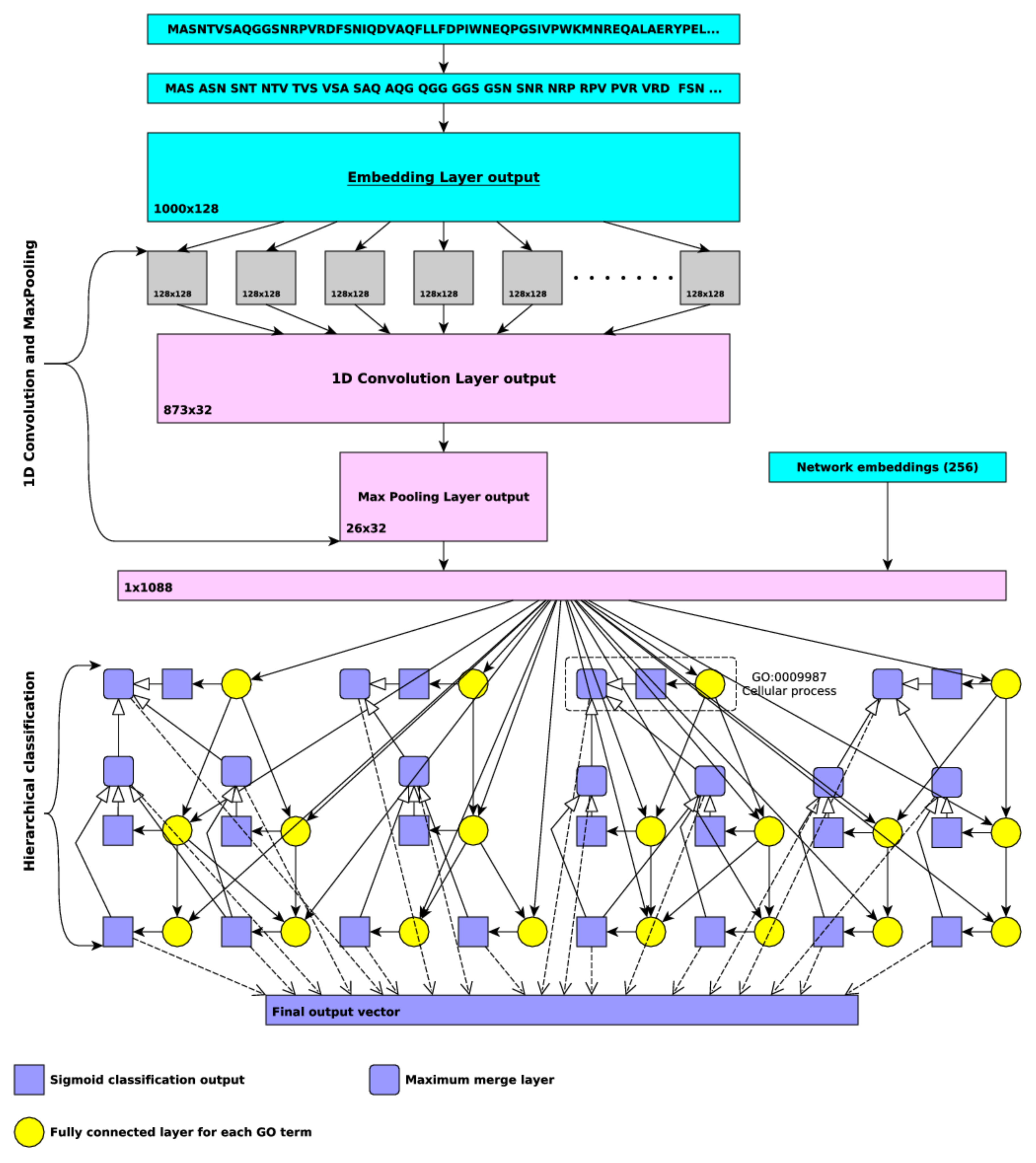}
\caption{\label{fig:1} Convolutional Neural Network Architecture. 1)
  The input of the model is a list of integer indexes of trigrams
  generated from protein sequence and vector of size 256 for protein
  PPI network representation. The trigram indexes are passed to
  embedding layer which provides vector representations of size 128
  for each trigram. The output of an embedding layer is a matrix of
  size 1000x128 on which we apply convolution and max-pooling. We
  merge the flattened output of the max-pooling layer with input PPI
  network representation which is then passed to hierarchical
  classification layers. 2) Hierarchical classification layers has a
  DAG structure of GO for is-a relations and for each GO term we created
  two fully-connected layers. First layer (yellow circle) has 256 neurons and 
  it takes protein representation. Second layer (purple square) has 1 neuron
  and it is connected to the first layers and performs the classification
  for one GO term. Additionally, internal nodes in the graph have maximum
  merge layers (rounded purple square) which outputs the maximum value of
  all the children and second layer of the term. Finally, the output
  vector of the model is the concatenation of maximum merge layers of the
  internal nodes and classification layers of the leaf nodes.}
\end{figure*}

Using a fully-connected layers for each class in GO, we created a
hierarchical classification neural network model. We use only the
subclass relations and create a small neural network for each class in
our subset of selected terms. Each network consists of two
fully-connected layers. The first layer has an output of 256 neurons
with a Rectified linear unit (RelU) activation function, and takes as
an input the protein representation concatenated with a first layer
outputs of the parent terms. The second layer has an output of a
single neuron with a sigmoid activation function and takes as an input
the output of the first layer. This layer is responsible for
classifying the proteins for its term. To ensure consistent
hierarchical classification, for each class which has children in GO,
we created a merge layer which selects the maximum value of the
classification layers of the term and its children. Finally, the
output of the model is the concatenation of classification layers of
leaf nodes and the maximum layers of internal nodes. Figure
\ref{fig:1} shows the architecture of our neural network model.

\subsection{Model implementation and optimization}
In training, we minimize the multi-output binary cross entropy loss
function using the Rmsprop optimizer \cite{Tieleman2012} with a mini
batch of size $128$ and learning rate of $0.01$. Initially, the
weights of our model are initialized according to a uniform
distribution \cite{glorot2010understanding}. We fit our model with
80\% of our training set and use the remaining 20\% of the training
set as a validation set. At the end of each training epoch, we monitor
the convergence of the model on the validation set and keep the
weights of the best performing model. To prevent over-fitting of the
model, we use dropout layers as regularizers. We implement our model
using the deep learning library Keras with TensorFlow
\cite{tensorflow2016} as a backend. To accelerate the training
process, we use NVIDIA Pascal X GPUs. We manually tuned the following
set of parameters: minibatch size, number of convolution filters,
filter size, number of neurons in fully connected layer, and learning
rate. Source code for our implementation is available at
\url{https://github.com/bio-ontology-research-group/deepgo}.

\subsection{BLAST baseline}
We use the BLAST \cite{blast1997} sequence alignment method as a
baseline to compare our model's performance. We use BLAST to find the
most similar sequence in a database of experimentally annotated
proteins for a query sequence and assign all its annotations to the
query sequence. We create a database for each ontology with a proteins
in our training set that have been annotated with at least one term
from the ontology. For a proteins in our test set, we use the BLASTP
program to obtain the protein with the highest alignment score from
our training set and assign all its functional terms to the protein
from our test set.

\subsection{Evaluation}

We evaluate our model performance with two measures \cite{Wyatt2013}
that are used in CAFA challenge \cite{cafa}. The first measure is a
protein centric maximum F-measure. Here, we compute F-measure for a
threshold $t \in [0, 1]$ using the average precision for proteins for
which we predict at least one term and average recall for all
proteins. Then, we select the maximum F-measure of all thresholds. We
compute the $F_{max}$ measure using the following formulas:
\begin{equation}
pr_{i}(t) = \frac{\sum{_f I(f \in P_i(t) \wedge f \in T_i)}}{\sum{_f I(f \in P_i(t))}}
\end{equation}
\begin{equation}
rc_{i}(t) = \frac{\sum{_f I(f \in P_i(t) \wedge f \in T_i)}}{\sum{_f I(f \in T_i)}}
\end{equation}
\begin{equation}
AvgPr(t) = \frac{1}{m(t)}*\sum_{i=1}^{m(t)}pr_{i}(t) 
\end{equation}
\begin{equation}
AvgRc(t) = \frac{1}{n}*\sum_{i=1}^{n}rc_{i}(t) 
\end{equation}
\begin{equation}
F_{max} = \max_{t}\left\{\frac{2 * AvgPr(t) * AvgRc(t)}{AvgPr(t) + AvgRc(t)}\right\}
\end{equation}
In these measures, $f$ is GO class, $P_i(t)$ is a set of predicted
classes for a protein $i$ using a threshold $t$, and $T_i$ is a set of
annotated classes for a protein $i$. Precision is averaged over the
proteins where we at least predict one term and $m(t)$ is the total
number of such proteins. $n$ is a number of all proteins in a test
set.

The second measure is a term-centric where for each term $f$ we
compute AUC of a ROC Curve of a sensitivity (or a recall) for a given
false positive rate (1 - specificity). We compute sensitivity and
specificity using the following formulas:
\begin{equation}
sn_{f}(t) = \frac{\sum{_i I(f \in P_i(t) \wedge f \in T_i)}}{\sum{_i I(f \in T_i)}}
\end{equation}

\begin{equation}
sp_{f}(t) = \frac{\sum{_i I(f \notin P_i(t) \wedge f \notin T_i)}}{\sum{_i I(f \notin T)}}
\end{equation}
Here, $P_i(t)$ is a set of predicted terms for a protein $i$ using a
threshold $t$ and $T_i$ is a set of annotated terms for a protein $i$.
Additionally, we report a term-centric $F_{max}$ measure where for
each term $f$ we compute the F-measure using threshold $t$ and all
proteins in our test set. Then, we take the maximum for all the
thresholds.
\begin{equation}
pr_{f}(t) = \frac{\sum{_i I(f \in P_i(t) \wedge f \in T_i)}}{\sum{_i I(f \in P_i(t))}}
\end{equation}
\begin{equation}
rc_{f}(t) = \frac{\sum{_i I(f \in P_i(t) \wedge f \in T_i)}}{\sum{_i I(f \in T_i)}}
\end{equation}
\begin{equation}
F_{max f} = \max_{t}\left\{\frac{2 * pr_{f}(t) * rc_{f}(t)}{pr_{f}(t) + rc_{f}(t)}\right\}
\end{equation}

\section{Results}

\subsection{Feature learning and neuro-symbolic hierarchical
  classification}
We build a machine learning model that aims to address three
challenges in computational function prediction: learning features to
represent a protein, predicting functions in a hierarchical output
space with strong dependencies, and combining information from protein
sequences with protein-protein interaction networks.
%
%
The first part of our model learns a vector representation for a
protein sequence which can be used as features to predict protein
functions.  The second part of the model aims to encode for the
functional dependencies between classes in GO and optimizes
classification accuracy over the hierarchical structure of GO at once
instead of optimizing one model locally for each class. The intention
is that this model can identify both explicit dependencies between
classes in GO, as expressed by relations between classes encoded
in the ontology, as well as implicit dependencies such as frequently
co-occurring classes. While a single model over the entire GO would
likely yield best results, due to the size of the GO, we independently
train three models for each of GO's three sub-ontologies, Molecular
Function (MF), Biological Process (BP), and Cellular Component (CC),
and focus exclusively on subclass relations between GO classes.  We
generate a series of fully connected layers, one for each class $C$ in
the GO. Each of these layers has exactly one connection to an output
neuron, $Out(C)$, and, for each direct subclass $D$ of $C$, a
connection to another layer representing $D$. This architecture
resembles the hierarchical structure of GO and the dependencies
between its classes, ensures that discriminating features of each
class can be learned hierarchically while taking into account the
symbolic relations in GO. More generally, each dense layer of this
ontology-structured neural network layout is intended to learn
features that can discriminate between its subclasses and will pass
these features on to the next layers. Figure \ref{fig:1} illustrates
the basic architecture of our model.

We train three model in a supervised way (one model for each of the GO
ontologies). For this purpose, we first split all proteins with
manually curated GO annotations in SwissProt in a training set (80\%)
and an evaluation set (20\%). We use the manually assigned GO
functions of the proteins in the training set to train our models.
The performance of each model is globally optimized over all the GO
functions (within either the MF, BP, or CC hierarchy) through
back-propagation.  We then evaluate the performance of our model on
the 20\% of proteins not used for training, using the evaluation
metrics developed and employed in the CAFA challenge
\cite{cafa}. Table \ref{tab:models} shows the overall performance of
our model and the comparison to using BLAST to assign functions. We
find that our model, which relies only on protein sequences
(DeepGOSeq), outperforms BLAST in predicting cellular locations, but
does not achieve improved performance compared to BLAST in the MF and
BP ontologies when evaluated either on the full set of GO functions or
the subset used by our model.


\subsection{Incorporating protein networks}

The majority of functions and biological processes in GO require
multiple proteins to be performed. One source of information for
proteins acting together can be obtained from protein-protein
interaction networks. By adding information about protein-protein
interactions, we planned to improve our model's performance, in
particular for prediction of associations to biological processes
which usually require more than one protein to be performed. We encode
protein-protein interactions as a multi-species knowledge graph of
interacting proteins in which proteins within a species are linked
through {\em interacts-with} edges and proteins in different species
through a {\em orthologous-to} edge. We then apply a method to
generate knowledge graph embeddings \cite{alshahrani16} to this graph
and generate a vector representation for each protein. Furthermore, we
integrate this vector representation with the protein sequence
representation in our model, resulting in a multi-modal model that
utilizes both protein sequences and protein interactions.
Incorporating this network information significantly improves the
performance for almost all GO classes, and the overall performance of
our DeepGO method improves significantly in comparison with DeepGOSeq
which uses only protein sequence as a feature, and in comparison to
the BLAST baseline. Table \ref{tab:models} summarizes the results.

We find that the predictive performance of our model varies
significantly between proteins in different organisms, in particular
between single-cell and multi-cellular organisms. Table
\ref{tab:organism} summarizes the performance we achieve for
individual organisms, and further broadly distinguishes between
eukaryotic and prokaryotic organisms. We find that DeepGO achieves
high performance for well-characterized model organisms, likely due
to the rich characterization of protein functions in these organisms;
other organisms do not have a large set of manually asserted function
annotations and are therefore represented more sparsely in our
evaluation set.

We further evaluated how well DeepGO performs on different types of
proteins. InterPro classifies proteins into families, domains and
important sites \cite{interpro2017}. We evaluate DeepGO's performance
by grouping proteins by their InterPro annotations. Supplementary
Table \ref{tab:interpro} shows the performance for InterPro classes with at least 50
protein annotations in our test set. We find that for some important
protein families, such as p53-like transcription factors ({\tt
  IPR008967}), DeepGO can achieve high performance in all three GO
ontologies, while for other kinds of proteins, such as those with a
Ubiquitin-related domain ({\tt IPR029071}), DeepGO fails to predict
annotations to BP and MF accurately.

\begin{figure*}[!ht]
  \centering
\includegraphics[width=1.0\textwidth]{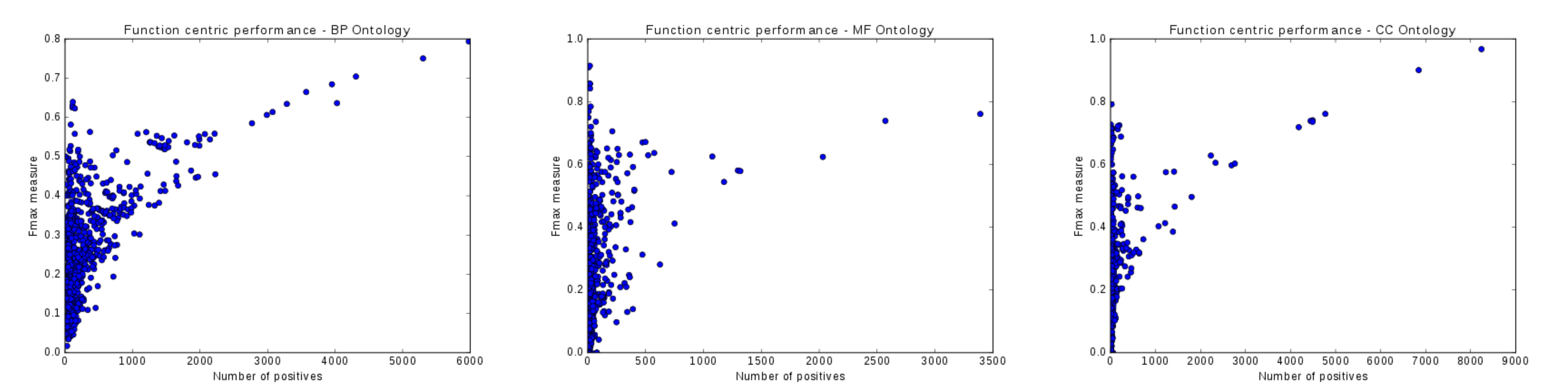}
\caption{\label{fig:2} Term centric performance. These plots show the performance of our model for each term in our subset of GO as a function of the number of supporting proteins in test set which are annotated by the term. }
\end{figure*}

We further perform a term-centric evaluation \cite{cafa} in which we
test how accurate our predictions are for different GO
functions. Supplementary Table \ref{tab:funcs} shows the best performing GO
functions from each ontology. Unsurprisingly, high-level functions
with a large number of annotations generally perform significantly
better than more specific functions.  We further test whether the
variance in predictive performance is intrinsic to our method or the
result of different amounts of training data available for proteins of
different families, with different domains, or for GO functions with
different number of annotations. We plot the predictive performance of
DeepGO as a function of the number of training samples in
Figure \ref{fig:2}, and observe that performance is strongly
correlated with the number of training instances. However, due to the
hierarchical nature of GO, an increased number of training instances
will always be available for more general, high-level functions.  In
the future, additional weights based on information content of GO
classes \cite{Resnik1999} should be assigned to more specific
functions which contain more information \cite{cafa, Wyatt2013}; using
these weights during training of our model may improve performance for
more specific functions.

\begin{table*}[ht]
\begin{adjustwidth}{-1.5in}{0in} 
\begin{tabular}{l l l l | l l l | l l l}
\textbf{Method} & \multicolumn{3}{l}{\textbf{BP}} & \multicolumn{3}{l}{\textbf{MF}} & \multicolumn{3}{l}{\textbf{CC}} \\
 & $\mathbf{F_{max}}$ & \textbf{AvgPr} & \textbf{AvgRc} & $\mathbf{F_{max}}$ & \textbf{AvgPr} & \textbf{AvgRc} & $\mathbf{F_{max}}$ & \textbf{AvgPr} & \textbf{AvgRc}\\
\toprule
BLAST & 0.31 & 0.30 & 0.33 & 0.37 & 0.37 & 0.38 & 0.36 & 0.32 & 0.42 \\
DeepGOSeq & 0.25 & 0.20 & 0.33 & 0.36 & 0.47 & 0.29 & 0.57 & 0.59 & 0.55 \\
DeepGO & \textbf{0.36} & \textbf{0.39} & \textbf{0.34} & \textbf{0.46} & \textbf{0.60} & \textbf{0.38} & \textbf{0.63} & \textbf{0.66} & \textbf{0.61} \\
\hline
BLAST (selected terms) & 0.34 & 0.38 & 0.32 & \textbf{0.54} & 0.61 & \textbf{0.48} & 0.50 & 0.51 & 0.49 \\
DeepGOSeq (selected terms) & 0.27 & 0.20 & 0.38 & 0.38 & 0.47 & 0.32 & 0.57 & 0.61 & 0.55 \\
DeepGO (selected terms) & \textbf{0.40} & \textbf{0.45} & \textbf{0.36} & 0.50 & \textbf{0.62} & 0.42 & \textbf{0.64} & \textbf{0.66} & \textbf{0.62} \\
\end{tabular}
\caption{\label{tab:models} Overview of our model's performance and
  comparison to BLAST baseline. The DeepGOSeq model uses only sequence
  information, while DeepGO uses both the protein sequence and network
  interactions as input. The first part of the evaluation shows
  performance results when considering all GO annotations (even those
  that our model cannot predict), while the second part focuses on the
  selected terms for which our model can generate predictions.}
\end{adjustwidth}
\end{table*}

\begin{table*}[ht]
\begin{adjustwidth}{-1.5in}{0in} 
\begin{tabular}{l l l l | l l l | l l l}
\textbf{Organism} & \multicolumn{3}{l}{\textbf{BP}} & \multicolumn{3}{l}{\textbf{MF}} & \multicolumn{3}{l}{\textbf{CC}} \\
 & $\mathbf{F_{max}}$ & \textbf{AvgPr} & \textbf{AvgRc} & $\mathbf{F_{max}}$ & \textbf{AvgPr} & \textbf{AvgRc} & $\mathbf{F_{max}}$ & \textbf{AvgPr} & \textbf{AvgRc}\\
\toprule
\textbf{Eukaryotes} & 0.36 & 0.39 & 0.34 & 0.48 & 0.63 & 0.39 & 0.63 & 0.65 & 0.62 \\
Human & 0.39 & 0.45 & 0.34 & 0.51 & 0.65 & 0.42 & 0.61 & 0.59 & 0.63 \\
Mouse & 0.37 & 0.40 & 0.34 & 0.49 & 0.61 & 0.41 & 0.60 & 0.65 & 0.56 \\
Rat & 0.36 & 0.37 & 0.35 & 0.50 & 0.63 & 0.42 & 0.55 & 0.56 & 0.54 \\
Fruit Fly & 0.36 & 0.42 & 0.32 & 0.50 & 0.62 & 0.43 & 0.57 & 0.66 & 0.49 \\
Yeast & 0.40 & 0.48 & 0.34 & 0.43 & 0.49 & 0.38 & 0.58 & 0.62 & 0.54 \\
Fission Yeast & 0.38 & 0.40 & 0.37 & 0.44 & 0.59 & 0.34 & 0.77 & 0.77 & 0.77 \\
Zebrafish & 0.37 & 0.45 & 0.31 & 0.56 & 0.59 & 0.53 & 0.66 & 0.65 & 0.67 \\
\hline
\textbf{Prokaryotes} & 0.36 & 0.41 & 0.32 & 0.38 & 0.45 & 0.33 & 0.66 & 0.75 & 0.60 \\
Ecoli & 0.37 & 0.42 & 0.32 & 0.38 & 0.49 & 0.31 & 0.68 & 0.79 & 0.60 \\
Mycobacterium tuberculosis & 0.37 & 0.51 & 0.29 & 0.39 & 0.51 & 0.32 & 0.69 & 0.75 & 0.63 \\
Pseudomonas aeruginosa & 0.46 & 0.57 & 0.47 & 0.31 & 0.32 & 0.37 & 0.50 & 1.00 & 0.33 \\
Bacillus subtilis & 0.38 & 0.46 & 0.32 & 0.43 & 0.42 & 0.44 & 0.43 & 1.00 & 0.28 \\
\end{tabular}
\caption{\label{tab:organism} Performance of our method distinguished
  by organisms. We use the DeepGO model that combines both sequence
  and network information for this prediction.}
\end{adjustwidth}
\end{table*}

\section{Discussion}

\subsection{Multi-modal function prediction}

Computational approaches to function prediction have been developed
for many years \cite{cafa}. One of the most basic approaches for
function prediction has been the use of BLAST \cite{blast1997} to
identify proteins with high sequence similarity and known functions,
and assign the functions of the best matching protein to the protein
to be characterized. Approaches for orthology-based function
prediction include more comprehensive modelling of evolutionary
relations, including relations between protein subdomains
\cite{paint}, and these can outperform simple BLAST baseline
experiments.  Other approaches for function prediction rely on
structure prediction. It is well known that protein ternary structure
strongly influences a protein's functions, but prediction of protein
structure remains a challenging computational problem \cite{casp}, and
even with known protein structure, functions cannot always be
predicted accurately. Additionally, high-level physiological
functions, such as {\em vocalization behavior} ({\tt GO:0071625}),
will not be predictable from a single protein's sequence or structure
alone but require complex pathways and interacting proteins, all of
which contribute to the function. 

While many of these approaches rely on hand-crafted features, some
approaches already applied feature learning (i.e., deep learning) to
parts of these data types. For example, feature learning approaches
have significantly improved the prediction of transcription factor
binding sites and functional impact of genomic variants \cite{deepsea,
  deepbind}. Here, we have extended the application of deep learning
approaches in function prediction in three ways: first, we apply
feature learning through the use of a CNN and embedding layer to
learn a representation of protein sequence; second, we developed a
deep, ontology-structured classification model that can refine
features on each distinction present in the GO; and third, we use
multi-modal data sources, in particular the protein structure and
information from protein-protein interaction networks, within a single
model. Through the multi-modal nature of our machine learning model,
other types of data can be integrated within the DeepGO model as long
as they can be used as input to a representation learning method that
learns vector representations. For example, protein structure
information, if available, could be incorporated in our model by
adding another feature learning branch that generates dense,
low-dimensional representations of protein structure \cite{Wang2017}
and using these as input to our hierarchical classifier.

\subsection{Hierarchical classification on ontologies}

In addition to the multi-modal nature of features used in DeepGO,
another contribution of our work is the deep hierarchical
classification model that optimizes predictive performance on whole
hierarchies, accounts for class dependencies, learns features in a
hierarchical manner, and is optimized jointly together with the
feature learning component of our model in an end-to-end manner. Our
method can be applied to other applications with a similarly
structured output space and which rely on learning feature
representations. In particular, we plan to apply our model for
predicting disease associations of genes which are encoded using the
Disease Ontology \cite{Osborne2009}, or phenotype associations of
genetic variants which are encoded using phenotype ontologies
\cite{pato-paper}.

The advantages of our model are its potential for end-to-end learning,
the global optimization, and the potential to predict any class given
sufficient training data. In particular the end-to-end learning
provides benefits over approaches such as structured support vector
machines \cite{SokolovB10}, which generally rely on hand-crafted
feature vectors. 

However, our model also has disadvantages. First, it needs large
amounts of training data for each class; this data is readily
available through the manual GO annotations that have been created for
many years, but will not easily be available for other areas of
application, such as predicting phenotype annotations or effects of
variants. Furthermore, our model is complex and requires large
computational resources for training, and therefore may not be
applicable in all settings.

In the future, we intend to extend our hierarchical model in several
directions. First, we plan to include more information from GO, in
particular parthood relations and regulatory relations, which may
provide additional information. We will also explore adding more
features, such as additional types of interactions (genetic
interactions, or co-expression networks), and information extracted
from text.


\clearpage

\setcounter{table}{0}

\section*{Supplementary materials}
\begin{table*}[ht]
\resizebox{1.0\textwidth}{!}{%
\begin{tabular}{l l l l l | l l l | l l l}
\textbf{InterPro} & \textbf{InterPro Name} & \multicolumn{3}{l}{\textbf{BP}} & \multicolumn{3}{l}{\textbf{MF}} & \multicolumn{3}{l}{\textbf{CC}} \\
 & & $\mathbf{F_{max}}$ & \textbf{AvgPr} & \textbf{AvgRc} & $\mathbf{F_{max}}$ & \textbf{AvgPr} & \textbf{AvgRc} & $\mathbf{F_{max}}$ & \textbf{AvgPr} & \textbf{AvgRc}\\
\toprule
IPR008967 & p53-like transcription factor, DNA-binding & 0.44 & 0.48 & 0.40 & 0.63 & 0.65 & 0.61 & 0.80 & 0.78 & 0.81 \\
IPR013083 & Zinc finger, RING/FYVE/PHD-type & 0.37 & 0.38 & 0.36 & 0.50 & 0.57 & 0.45 & 0.67 & 0.66 & 0.68 \\
IPR017907 & Zinc finger, RING-type, conserved site & 0.26 & 0.33 & 0.22 & 0.39 & 0.61 & 0.28 & 0.57 & 0.54 & 0.60 \\
IPR013087 & Zinc finger C2H2-type & 0.47 & 0.44 & 0.51 & 0.58 & 0.53 & 0.64 & 0.77 & 0.75 & 0.79 \\
IPR011991 & Winged helix-turn-helix DNA-binding domain & 0.39 & 0.41 & 0.37 & 0.41 & 0.52 & 0.34 & 0.64 & 0.65 & 0.64 \\
IPR015943 & WD40/YVTN repeat-like-containing domain & 0.41 & 0.41 & 0.40 & 0.52 & 0.62 & 0.45 & 0.66 & 0.66 & 0.65 \\
IPR019775 & WD40 repeat, conserved site & 0.40 & 0.39 & 0.41 & 0.55 & 0.60 & 0.50 & 0.67 & 0.67 & 0.68 \\
IPR001680 & WD40 repeat & 0.41 & 0.41 & 0.40 & 0.53 & 0.60 & 0.48 & 0.67 & 0.68 & 0.67 \\
IPR029071 & Ubiquitin-related domain & 0.28 & 0.28 & 0.28 & 0.48 & 0.54 & 0.43 & 0.63 & 0.67 & 0.61 \\
IPR016135 & Ubiquitin-conjugating enzyme/RWD-like & 0.42 & 0.39 & 0.44 & 0.57 & 0.58 & 0.57 & 0.65 & 0.62 & 0.69 \\
IPR023313 & Ubiquitin-conjugating enzyme, active site & 0.42 & 0.39 & 0.46 & 0.59 & 0.57 & 0.61 & 0.65 & 0.59 & 0.71 \\
IPR018200 & Ubiquitin specific protease, conserved site & 0.31 & 0.32 & 0.31 & 0.46 & 0.53 & 0.40 & 0.67 & 0.67 & 0.67 \\
IPR028889 & Ubiquitin specific protease domain & 0.32 & 0.32 & 0.31 & 0.46 & 0.53 & 0.40 & 0.67 & 0.67 & 0.67 \\
IPR012336 & Thioredoxin-like fold & 0.26 & 0.28 & 0.25 & 0.48 & 0.50 & 0.47 & 0.66 & 0.64 & 0.68 \\
IPR000727 & Target SNARE coiled-coil homology domain & 0.44 & 0.46 & 0.42 & 0.56 & 0.81 & 0.43 & 0.53 & 0.73 & 0.41 \\
IPR008271 & Serine/threonine-protein kinase, active site & 0.41 & 0.42 & 0.40 & 0.63 & 0.70 & 0.57 & 0.59 & 0.61 & 0.58 \\
IPR001452 & SH3 domain & 0.35 & 0.40 & 0.31 & 0.44 & 0.55 & 0.36 & 0.55 & 0.56 & 0.54 \\
IPR000980 & SH2 domain & 0.40 & 0.43 & 0.37 & 0.53 & 0.66 & 0.44 & 0.65 & 0.66 & 0.65 \\
IPR029063 & S-adenosyl-L-methionine-dependent methyltransferase & 0.43 & 0.43 & 0.43 & 0.43 & 0.57 & 0.35 & 0.73 & 0.71 & 0.75 \\
IPR000504 & RNA recognition motif domain & 0.45 & 0.46 & 0.45 & 0.70 & 0.66 & 0.74 & 0.69 & 0.66 & 0.72 \\
IPR011009 & Protein kinase-like domain & 0.39 & 0.42 & 0.36 & 0.61 & 0.67 & 0.56 & 0.59 & 0.61 & 0.57 \\
IPR017441 & Protein kinase, ATP binding site & 0.41 & 0.45 & 0.38 & 0.64 & 0.70 & 0.60 & 0.57 & 0.59 & 0.56 \\
IPR011993 & PH domain-like & 0.38 & 0.42 & 0.34 & 0.45 & 0.61 & 0.36 & 0.55 & 0.54 & 0.57 \\
IPR027417 & P-loop containing nucleoside triphosphate hydrolase & 0.37 & 0.40 & 0.34 & 0.37 & 0.60 & 0.27 & 0.63 & 0.66 & 0.61 \\
IPR016040 & NAD(P)-binding domain & 0.37 & 0.39 & 0.35 & 0.40 & 0.59 & 0.30 & 0.73 & 0.70 & 0.75 \\
IPR020846 & Major facilitator superfamily domain & 0.44 & 0.47 & 0.41 & 0.43 & 0.65 & 0.32 & 0.61 & 0.66 & 0.57 \\
IPR032675 & Leucine-rich repeat domain, L domain-like & 0.32 & 0.45 & 0.25 & 0.47 & 0.70 & 0.36 & 0.48 & 0.55 & 0.43 \\
IPR013783 & Immunoglobulin-like fold & 0.33 & 0.36 & 0.31 & 0.44 & 0.61 & 0.34 & 0.49 & 0.53 & 0.46 \\
IPR009057 & Homeobox domain-like & 0.39 & 0.44 & 0.35 & 0.57 & 0.64 & 0.52 & 0.75 & 0.74 & 0.75 \\
IPR009071 & High mobility group box domain & 0.40 & 0.41 & 0.38 & 0.64 & 0.62 & 0.66 & 0.74 & 0.72 & 0.75 \\
IPR011992 & EF-hand domain pair & 0.35 & 0.45 & 0.28 & 0.53 & 0.62 & 0.46 & 0.63 & 0.70 & 0.57 \\
IPR013320 & Concanavalin A-like lectin/glucanase domain & 0.26 & 0.34 & 0.21 & 0.35 & 0.60 & 0.25 & 0.57 & 0.56 & 0.57 \\
IPR000008 & C2 domain & 0.33 & 0.39 & 0.29 & 0.38 & 0.54 & 0.29 & 0.50 & 0.69 & 0.39 \\
\end{tabular}
}
\caption{\label{tab:interpro} Performance of DeepGO by InterPro domains. Only InterPro
  domains for which at least 50 proteins are in our evaluation dataset
  are included in this evaluation.}
\end{table*}

\begin{table*}[ht]
\resizebox{1.0\textwidth}{!}{%
\begin{tabular}{l l l l l l}
Function & Label & \multicolumn{2}{l}{\textbf{DeepGO}} & \multicolumn{2}{l}{\textbf{DeepGOSeq}} \\
 & & $\mathbf{F_{max}}$ & \textbf{ROC AUC} & $\mathbf{F_{max}}$ & \textbf{ROC AUC} \\
\toprule
\textbf{Biological Process} & & & & & \\
GO:0009987 & cellular process & 0.793545 & 0.680064 & 0.793765 & 0.540680 \\
GO:0044699 & single-organism process & 0.750229 & 0.699637 & 0.738709 & 0.577044 \\
GO:0065007 & biological regulation & 0.704066 & 0.759365 & 0.677123 & 0.689570 \\
GO:0008152 & metabolic process & 0.634190 & 0.759272 & 0.544373 & 0.608695 \\
GO:0032502 & developmental process & 0.551414 & 0.620468 & 0.392551 & 0.625560 \\
GO:0050896 & response to stimulus & 0.454906 & 0.683399 & 0.393794 & 0.512953 \\
GO:0071840 & cellular component organization or biogenesis & 0.448368 & 0.703925 & 0.365183 & 0.572168 \\
GO:0051179 & localization & 0.426162 & 0.708099 & 0.311107 & 0.489611 \\
GO:0032501 & multicellular organismal process & 0.413983 & 0.531594 & 0.223127 & 0.566755 \\
GO:0040007 & growth & 0.403571 & 0.237761 & 0.074300 & 0.326924 \\
GO:0002376 & immune system process & 0.383117 & 0.337897 & 0.085541 & 0.393173 \\
GO:0022414 & reproductive process & 0.370014 & 0.434946 & 0.190107 & 0.566350 \\
GO:0051704 & multi-organism process & 0.277030 & 0.512176 & 0.169651 & 0.537286 \\
GO:0007610 & behavior & 0.262774 & 0.217460 & 0.049016 & 0.405271 \\
GO:0040011 & locomotion & 0.200238 & 0.415350 & 0.071258 & 0.537358 \\
GO:0022610 & biological adhesion & 0.153846 & 0.145779 & 0.042748 & 0.187945 \\
GO:0023052 & signaling & 0.150171 & 0.084057 & 0.017836 & 0.010441 \\
GO:0048511 & rhythmic process & 0.116883 & 0.057025 & 0.015441 & 0.027048 \\
GO:0000003 & reproduction & 0.072398 & 0.010535 & 0.013423 & 0.002335 \\
\midrule
\textbf{Molecular Function} & & & & & \\
GO:0005488 & binding & 0.760884 & 0.778792 & 0.726436 & 0.714915 \\
GO:0003824 & catalytic activity & 0.738823 & 0.835322 & 0.671065 & 0.732225 \\
GO:0005215 & transporter activity & 0.636451 & 0.314824 & 0.594164 & 0.450674 \\
GO:0001071 & nucleic acid binding transcription factor activity & 0.519453 & 0.362382 & 0.388293 & 0.448391 \\
GO:0060089 & molecular transducer activity & 0.502392 & 0.350086 & 0.342723 & 0.384180 \\
GO:0004871 & signal transducer activity & 0.481572 & 0.378499 & 0.343465 & 0.496081 \\
GO:0098772 & molecular function regulator & 0.329268 & 0.334650 & 0.179612 & 0.576424 \\
GO:0016209 & antioxidant activity & 0.325926 & 0.062499 & 0.056395 & 0.025002 \\
GO:0000988 & transcription factor activity, protein binding & 0.293413 & 0.239255 & 0.176398 & 0.333591 \\
GO:0005198 & structural molecule activity & 0.242152 & 0.277404 & 0.058824 & 0.426995 \\
GO:0009055 & electron carrier activity & 0.187500 & 0.017467 & 0.027778 & 0.040924 \\
GO:0045182 & translation regulator activity & 0.070175 & 0.032282 & 0.007722 & 0.027473 \\
\midrule
\textbf{Cellular Component} & & & & & \\
GO:0044464 & cell part & 0.967330 & 0.826043 & 0.966631 & 0.697060 \\
GO:0043226 & organelle & 0.761161 & 0.595503 & 0.708719 & 0.645590 \\
GO:0016020 & membrane & 0.605258 & 0.691536 & 0.500599 & 0.710984 \\
GO:0044422 & organelle part & 0.602635 & 0.630250 & 0.495917 & 0.630139 \\
GO:0044421 & extracellular region part & 0.498270 & 0.306901 & 0.165513 & 0.575960 \\
GO:0032991 & macromolecular complex & 0.465488 & 0.653815 & 0.335831 & 0.638300 \\
GO:0005576 & extracellular region & 0.452276 & 0.248848 & 0.368515 & 0.542654 \\
GO:0044425 & membrane part & 0.402873 & 0.505403 & 0.301491 & 0.580220 \\
GO:0044456 & synapse part & 0.371429 & 0.084840 & 0.020779 & 0.004898 \\
GO:0099512 & supramolecular fiber & 0.345946 & 0.098424 & 0.078240 & 0.021825 \\
GO:0045202 & synapse & 0.309524 & 0.032163 & 0.000000 & 0.000000 \\
GO:0031974 & membrane-enclosed lumen & 0.303226 & 0.199096 & 0.044743 & 0.202145 \\
GO:0031012 & extracellular matrix & 0.291971 & 0.098603 & 0.012712 & 0.014834 \\
GO:0030054 & cell junction & 0.242424 & 0.243079 & 0.062822 & 0.305787 \\
GO:0009295 & nucleoid & 0.200000 & 0.003091 & 0.000000 & 0.000000 \\
GO:0044420 & extracellular matrix component & 0.125000 & 0.001101 & 0.000000 & 0.000000 \\
GO:0044217 & other organism part & 0.111940 & 0.047736 & 0.027149 & 0.004069 \\
GO:0005623 & cell & 0.068966 & 0.022208 & 0.018182 & 0.000568 \\
GO:0044423 & virion part & 0.066158 & 0.039468 & 0.000000 & 0.000000 \\
GO:0019012 & virion & 0.029412 & 0.019432 & 0.000000 & 0.000000 \\
\end{tabular}
}
\caption{\label{tab:funcs} Performance of DeepGO distinguished by GO functions.}
\end{table*}
\clearpage

\section*{Acknowledgements}
We acknowledge use of the compute resources of the Computational
Bioscience Research Center (CBRC) at King Abdullah University of
Science and Technology (KAUST).

\section*{Funding}
This work was supported by funding from King Abdullah University of
Science and Technology (KAUST) [FCC/1/1976-08-01].

\bibliography{lc}

\bibliographystyle{abbrv}

\end{document}